\documentclass[a4paper,fleqn]{cas-sc}

\usepackage{amsmath,amsfonts,amssymb}
\usepackage{booktabs}
\usepackage{listings}
\usepackage{xcolor}
\usepackage{microtype}
\usepackage{xurl}

\usepackage{tikz}
\usetikzlibrary{shapes.geometric,arrows.meta,positioning}

\newtheorem{theorem}{Theorem}
\newtheorem{lemma}[theorem]{Lemma}
\newtheorem{proposition}[theorem]{Proposition}
\newdefinition{definition}{Definition}
\newproof{proof}{Proof}

\begin{document}
\let\WriteBookmarks\relax
\def\floatpagepagefraction{1}
\def\textpagefraction{.001}

\shorttitle{Restoring Fair Hrtick Re-arming After Same-Task Repicks}
\shortauthors{Shubhang Kaushik}

\renewcommand{\eadauthor}{Shubhang Kaushik}

\title[mode = title]{Restoring Fair Hrtick Re-arming After Same-Task Repicks in the Linux EEVDF Scheduler}

\author[1]{Shubhang Kaushik}[role={Software Engineer, Linux Kernel Architecture and Scalability},
                        orcid=0000-0001-5590-1015]
\cormark[1]
\ead{shubhangpk@gmail.com}
\ead{sh@gentwo.org}
\credit{Conceptualization, Methodology, Software, Investigation, Validation, Formal Analysis, Data Curation, Writing - Original Draft, Writing - Review \& Editing}

\address[1]{Scalability and Performance Group, OS Enabling Division, Ampere Computing, Santa Clara, CA, USA}

\cortext[cor1]{Corresponding author. Email addresses: shubhangpk@gmail.com, sh@gentwo.org}

\begin{abstract}
Linux uses one-shot high-resolution scheduler ticks (hrticks) to request rescheduling at fair-class execution boundaries. We analyze a gap in the dispatch path in which an expired fair hrtick is not restarted when the fair scheduler selects the current task again. The gap arises because the same-task fast path bypasses \texttt{set\_next\_task\_fair()}, the location that normally records the next fair hrtick delay.

We introduce \texttt{SNT\_REPICK}, an explicit same-task activation type for \texttt{struct sched\_class}. For fair tasks, it preserves the fast path while restarting the hrtick from refreshed EEVDF state. For \texttt{SCHED\_DEADLINE}, it deliberately performs no hrtick restart because the corresponding runtime state is not refreshed on a repick. The implementation was merged into the mainline Linux kernel as commit \texttt{c72945693b}.

We give a conditional timing analysis for systems on which the periodic scheduler tick remains active and report validation from two platforms. Functional tests on an 80-core Ampere Altra exercised fair hrtick, deadline hrtick, and the \texttt{CONFIG\_HIGH\_RES\_TIMERS=n} fallback. Independently, Intel's 0-Day Kernel Test Service reported a 34.2\% increase in its \texttt{stress-ng.rawudp.packets\_32\_bytes\_received\_per\_sec} metric on a 192-thread Intel Xeon 6740E under the documented test configuration. We distinguish this reported benchmark result from causal claims that require separate scheduler and network instrumentation.
\end{abstract}


\maketitle

\section{Introduction}
Operating systems deployed across cloud datacenters and high-performance computing (HPC) nodes must arbitrate compute capacity among heterogeneous workloads. Batch simulations, networking daemons, and message-passing pipelines can share processor cores, making both proportional allocation and timely scheduling decisions important \cite{lozi2016cfs}. The CPU scheduler is the operating-system component that mediates these decisions \cite{corbet2023eevdf}.

For nearly seventeen years, the default general-purpose scheduler in the Linux kernel was the Completely Fair Scheduler (CFS) \cite{lozi2016cfs}. CFS approximated Generalized Processor Sharing (GPS) by balancing tasks across red-black trees indexed by virtual execution time ($vruntime$). However, CFS inherently conflated throughput allocation with dispatch latency, relying on an intricate web of ad-hoc heuristics (such as sleeper credits, wake-up preemption granularity tunables, and latency-sensitive burst mechanisms) to prevent interactive starvation \cite{lozi2016cfs}. In Linux version 6.6, the kernel transitioned its fair scheduling class to the Earliest Eligible Virtual Deadline First (EEVDF) scheduler \cite{zijlstra2023eevdf, stoica1996eevdf}. EEVDF formally decouples bandwidth allocation (governed by entity weights $w_i$) from dispatch latency (governed by request slices $q_i$ and virtual deadlines $d_i$).

In current Linux kernels, EEVDF supplies the selection policy within the fair scheduling class. Linux configurations using conventional periodic ticks inspect task state at a frequency selected by \texttt{CONFIG\_HZ}, commonly 100--1000 Hz \cite{gleixner2006hrtimers}. Where enabled, \texttt{CONFIG\_SCHED\_HRTICK} supplements this mechanism with one-shot high-resolution timer events. Hrticks request rescheduling at a calculated fair-scheduling boundary; they do not by themselves establish an end-to-end application-latency guarantee.

\subsection{Missed Fair-Hrtick Restart on a Same-Task Repick}
The implementation before commit \texttt{c72945693b} contained a long-standing activation gap in the core scheduler dispatch loop (\texttt{kernel/sched/core.c} and \texttt{kernel/sched/fair.c}):
\begin{enumerate}
    \item \textbf{One-Shot Timer Mechanics:} Hardware timers armed via \texttt{hrtick\_start\_fair()} are programmed as one-shot events. Upon expiration, the timer disarms, enters interrupt context, and marks the task for rescheduling by setting \texttt{TIF\_NEED\_RESCHED}.
    \item \textbf{Same-Task Selection in Saturated Runtimes:} During the invocation of \texttt{\_\_schedule()}, the scheduling class picks the next entity. In high-concurrency polling loops (e.g., multi-threaded raw UDP ingestion), the current task frequently remains the earliest eligible task with the minimum deadline.
    \item \textbf{The Fast-Path Dispatcher Bypass:} To minimize context-switch latency, \texttt{put\_prev\_set\_next\_task()} checks whether $next == prev$. If true, the legacy kernel executed an immediate return, bypassing the scheduling class method \texttt{set\_next\_task\_fair()}.
    \item \textbf{The Missed Restart:} Because \texttt{set\_next\_task\_fair()} was bypassed, no subsequent fair high-resolution timer was armed. The task returned to user space with its one-shot timer expired and cleared. On a configuration with an active periodic tick and absent another scheduling event, the next periodic tick is the next scheduler timing opportunity.
\end{enumerate}

The defect can defer the next fair hrtick after a same-task repick. Its application-level impact depends on workload, kernel configuration, tick behavior, and other rescheduling events; these factors motivate the scoped analysis and measurements in this paper.

\subsection{Contributions}
In this paper, we analyze and repair this missed restart:
\begin{itemize}
    \item \textbf{Conditional Timing Analysis:} We formulate the delay caused by a lost one-shot timer under explicit periodic-tick assumptions and identify why the bound does not generalize unchanged to tickless configurations.
    \item \textbf{The \texttt{SNT\_REPICK} Architectural Protocol:} We introduce a unified scheduler activation protocol into \texttt{struct sched\_class}, defining explicit semantics for normal activation (\texttt{SNT\_NORMAL}), cross-task switches (\texttt{SNT\_PICK}), and same-task repicks (\texttt{SNT\_REPICK}).
    \item \textbf{Cross-Class Safety:} We explain why a fair repick can restart its hrtick after the fair entity state is refreshed, whereas a deadline repick must not restart its hrtick from stale runtime state.
    \item \textbf{Evidence and Scope:} We document functional validation on Ampere Altra and a public 0-Day Kernel Test Service comparison that reports a 34.2\% improvement in one \texttt{stress-ng.rawudp} metric on Intel Xeon 6740E hardware \cite{lkp2026rawudp}.
\end{itemize}

\section{Background and Mathematical Formulation}

\subsection{Generalized Processor Sharing and EEVDF Mechanics}
Consider a system with $N$ active runnable entities $\mathcal{T} = \{T_1, T_2, \dots, T_N\}$. Each entity $T_i$ is characterized by a static or dynamic weight $w_i > 0$ and an allocated time slice quantum $q_i > 0$. The total runqueue weight is $W = \sum_{j=1}^N w_j$.

In an ideal Generalized Processor Sharing (GPS) model \cite{parekh1993gps}, the system virtual time $V(t)$ progresses inversely proportional to the aggregate weight:
\begin{equation}
V(t) = \int_0^t \frac{1}{W(\tau)} \, d\tau
\end{equation}
For any task $T_i$ running continuously on a physical processor over physical time interval $[t, t + \Delta t)$, its allocated virtual runtime advances as:
\begin{equation}
v_i(t + \Delta t) = v_i(t) + \frac{\Delta t}{w_i}
\end{equation}

EEVDF ensures proportional share allocation by maintaining the allocation lag \cite{stoica1996eevdf}:
\begin{equation}
lag_i(t) = V(t) - v_i(t)
\end{equation}
An entity is eligible to execute if and only if $lag_i(t) \ge 0 \iff v_i(t) \le V(t)$. When eligible, its priority is ordered by its virtual deadline $d_i(t)$:
\begin{equation}
d_i(t) = v_i(t) + \frac{q_i}{w_i}
\end{equation}
The scheduler selects the entity $T^*$ satisfying:
\begin{equation}
T^*(t) = \arg\min_{T_i \in \mathcal{T}, \, v_i(t) \le V(t)} \{d_i(t)\}
\end{equation}

\subsection{High-Resolution Preemption and a Conditional Disarming Bound}
When \texttt{CONFIG\_SCHED\_HRTICK} is enabled, the kernel records a delay for a high-resolution timer event at physical time:
\begin{equation}
t_{\text{preempt}} = t_{\text{current}} + \delta_{\text{slice}}
\end{equation}
where $\delta_{\text{slice}} = \min(q_i - \Delta t_{\text{consumed}}, \, \delta_{\text{quantum}})$. 

\begin{definition}[Timer-Event Delay]
Let $\epsilon_{\text{irq}}$ denote the delay from an hrtick expiry to the execution of its interrupt handler. This is a timer-event quantity, not an end-to-end application-latency bound:
\begin{equation}
\mathcal{L}_{\text{timer}} \le \epsilon_{\text{irq}}.
\end{equation}
\end{definition}

\begin{proposition}[Conditional delay after a lost fair hrtick]
Assume that (i) the periodic scheduler tick remains active with frequency $f_{\text{hz}}$, (ii) no earlier wakeup, interrupt, or explicit yield invokes scheduling, and (iii) the same task remains runnable. Let $\tau_{\text{jiffy}} = 1/f_{\text{hz}}$. If a same-task repick skips timer re-arming at physical time $t_0$, the time until the next periodic scheduling opportunity is bounded by:
\begin{equation}
0 < \Delta \mathcal{S} \leq \tau_{\text{jiffy}}.
\end{equation}
\end{proposition}

\begin{proof}
Under the stated assumptions, hrtick expiry leaves no newly armed fair hrtick after the fast-path return. The next periodic tick occurs no later than one tick period after $t_0$, giving the bound. The proposition does not apply to tickless operation, to CPUs for which the periodic tick is stopped, or when another event causes scheduling earlier.
\end{proof}

\begin{lemma}[Weighted delay model]
Let the runnable competitors of $T_i$ have aggregate weight $W-w_i$. In the simplified work-conserving model used here, a lost scheduling opportunity of duration $\Delta\mathcal{S}$ defers their aggregate proportional service by:
\begin{equation}
\mathcal{D}_{\text{queue}} = \frac{W - w_i}{W} \cdot \Delta \mathcal{S}.
\end{equation}
\end{lemma}

For example, with \texttt{CONFIG\_HZ=250}, $\tau_{\text{jiffy}}=4\,\text{ms}$. A lost hrtick can therefore defer the next periodic scheduling opportunity by nearly $4\,\text{ms}$ under the proposition's assumptions. This example is a scheduler-level timing bound; it does not by itself establish packet loss, throughput loss, or an application-latency bound.

\begin{figure}[pos=ht]
\centering
\begin{tikzpicture}[
  xscale=1.05,
  every node/.style={font=\sffamily\scriptsize}
]
\draw[thick, ->] (0, 2.5) -- (7.5, 2.5) node[right] {Time};
\node[anchor=west, font=\sffamily\bfseries\scriptsize] at (0, 3.0) {(a) Baseline Linux Kernel (Disarmed Timer Failure)};

\draw[fill=blue!15] (0.2, 2.2) rectangle (2.2, 2.8) node[pos=.5] {Quantum ($3\,\text{ms}$)};
\draw[fill=red!20] (2.2, 2.2) rectangle (6.2, 2.8) node[pos=.5] {\textbf{Unchecked Overrun ($+3.8\,\text{ms}$)}};

\draw[dashed, red, thick] (2.2, 2.0) -- (2.2, 3.0) node[above, font=\tiny\color{red}] {hrtick expires (one-shot)};
\node[below, font=\tiny, align=center] at (2.2, 1.8) {next == prev bypasses\\set\_next\_task\_fair()};
\draw[dashed, black!60] (6.2, 2.0) -- (6.2, 3.0) node[above, font=\tiny] {Coarse Jiffy Tick ($CONFIG\_HZ$)};

\draw[thick, ->] (0, 0.5) -- (7.5, 0.5) node[right] {Time};
\node[anchor=west, font=\sffamily\bfseries\scriptsize] at (0, 1.0) {(b) Patched with \texttt{SNT\_REPICK} (Continuous Enforcement)};

\draw[fill=teal!20] (0.2, 0.2) rectangle (2.2, 0.8) node[pos=.5] {Quantum 1 ($3\,\text{ms}$)};
\draw[fill=teal!30] (2.25, 0.2) rectangle (4.25, 0.8) node[pos=.5] {Quantum 2 ($3\,\text{ms}$)};
\draw[fill=teal!40] (4.30, 0.2) rectangle (6.30, 0.8) node[pos=.5] {Quantum 3 ($3\,\text{ms}$)};

\draw[dashed, teal!80!black, thick] (2.2, 0.0) -- (2.2, 1.0) node[above, font=\tiny\color{teal!80!black}] {repick: Timer Restarted};
\draw[dashed, teal!80!black, thick] (4.25, 0.0) -- (4.25, 1.0) node[above, font=\tiny\color{teal!80!black}] {repick: Timer Restarted};
\end{tikzpicture}
\caption{Illustrative timing profile under the active-periodic-tick assumptions in Section~2. In the baseline, skipping \texttt{set\_next\_task\_fair()} leaves no fair hrtick active. \texttt{SNT\_REPICK} records the next fair hrtick delay at every repick boundary.}\label{fig:timing}
\end{figure}
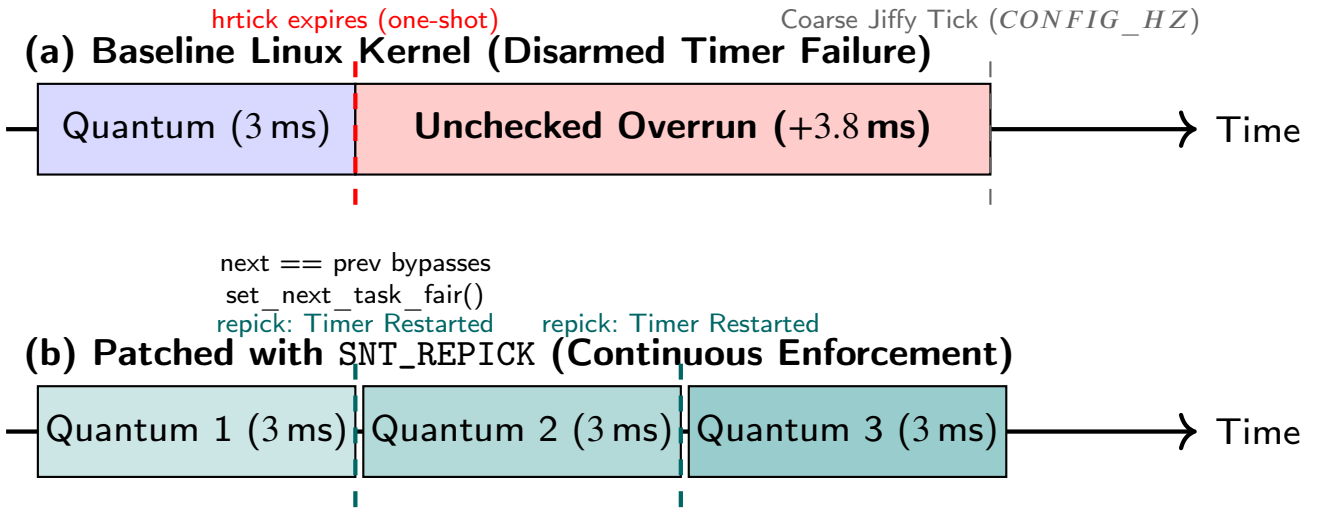

\section{The SNT\_REPICK Protocol Architecture}

To repair the missing activation while preserving the same-task fast path, the merged patch introduces the \textbf{Same-Next-Task Repick (\texttt{SNT\_REPICK})} activation type \cite{kaushik2026repick}.

\subsection{Formal State Machine}
Prior to our patch, the scheduling class interface exposed a boolean flag \texttt{first} to \texttt{set\_next\_task()}:
\begin{lstlisting}
void (*set_next_task)(struct rq *rq, struct task_struct *p, bool first);
\end{lstlisting}
This binary representation could not distinguish between a fresh task selection after a context switch and a same-task reselection within the fast-path. We replaced this flag across the kernel with an explicit enumeration:

\begin{lstlisting}
enum snt_e {
    SNT_NORMAL, /* External task activation via set_next_task() */
    SNT_PICK,   /* put_prev_set_next_task(): Standard transition (prev != next) */
    SNT_REPICK, /* put_prev_set_next_task(): Fast-path repick (prev == next) */
};
\end{lstlisting}

The following exact hunk from \texttt{kernel/sched/sched.h} shows the relevant dispatch change:

\begin{lstlisting}
static inline void put_prev_set_next_task(struct rq *rq,
                                          struct task_struct *prev,
                                          struct task_struct *next)
{
    __put_prev_set_next_dl_server(rq, prev, next);

    if (next == prev) {
        next->sched_class->set_next_task(rq, next, SNT_REPICK);
        return;
    }

    prev->sched_class->put_prev_task(rq, prev, next);
    next->sched_class->set_next_task(rq, next, SNT_PICK);
}
\end{lstlisting}

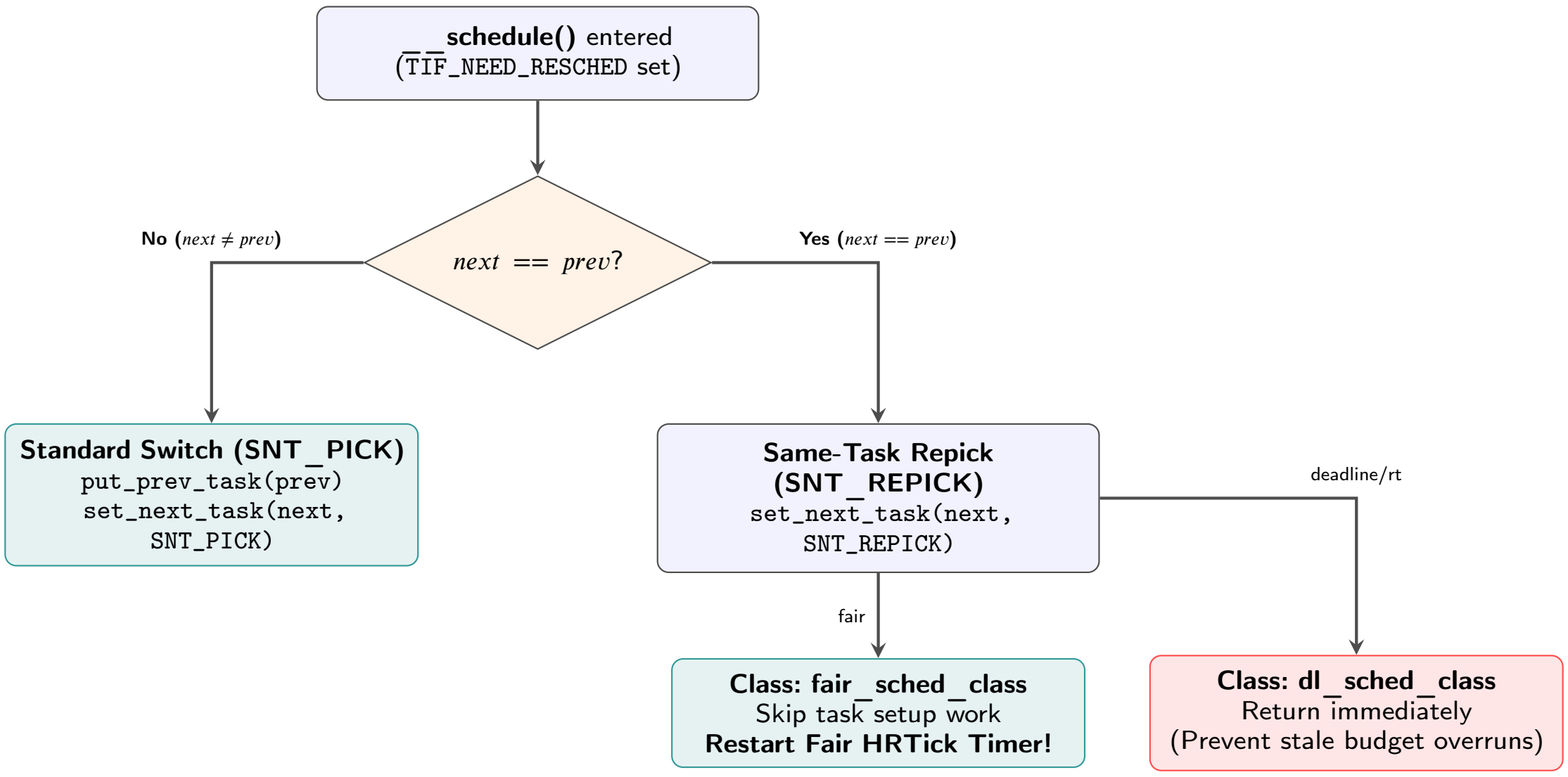
\begin{figure}[pos=ht]
\centering
\begin{tikzpicture}[
  node distance=1.1cm and 1.5cm,
  every node/.style={font=\sffamily\scriptsize},
  box/.style={rectangle, draw=black!70, fill=blue!5, rounded corners=3pt, text width=3.8cm, align=center, inner sep=5pt},
  decision/.style={diamond, draw=black!70, fill=orange!10, aspect=2, text width=2.4cm, align=center, inner sep=2pt},
  action/.style={rectangle, draw=teal!80, fill=teal!10, rounded corners=3pt, text width=3.6cm, align=center, inner sep=4pt},
  bypass/.style={rectangle, draw=red!70, fill=red!10, rounded corners=3pt, text width=3.6cm, align=center, inner sep=4pt},
  arr/.style={-stealth, thick, draw=black!70}
]
\node[box] (sched) {\textbf{\_\_schedule()} entered\\(\texttt{TIF\_NEED\_RESCHED} set)};
\node[decision, below=0.7cm of sched] (check) {$next == prev$?};
\node[action, below left=1.1cm and 0.3cm of check] (diff) {\textbf{Standard Switch (SNT\_PICK)}\\\texttt{put\_prev\_task(prev)}\\\texttt{set\_next\_task(next, SNT\_PICK)}};
\node[box, below right=1.1cm and 0.3cm of check] (same) {\textbf{Same-Task Repick (SNT\_REPICK)}\\\texttt{set\_next\_task(next, SNT\_REPICK)}};

\node[action, below=0.8cm of same] (fair) {\textbf{Class: fair\_sched\_class}\\Skip task setup work\\\textbf{Restart Fair HRTick Timer!}};
\node[bypass, right=0.6cm of fair] (dl) {\textbf{Class: dl\_sched\_class}\\Return immediately\\(Prevent stale budget overruns)};

\draw[arr] (sched) -- (check);
\draw[arr] (check) -| node[above, font=\tiny\bfseries] {No ($next \neq prev$)} (diff);
\draw[arr] (check) -| node[above, font=\tiny\bfseries] {Yes ($next == prev$)} (same);
\draw[arr] (same) -- node[left, font=\tiny] {fair} (fair);
\draw[arr] (same) -| node[above, font=\tiny] {deadline/rt} (dl);
\end{tikzpicture}
\caption{The \texttt{SNT\_REPICK} dispatcher state-machine transition in \texttt{put\_prev\_set\_next\_task()}. Fair tasks safely re-arm their one-shot hardware tick timers while real-time deadline tasks bypass re-arming to preserve execution budget invariants.}\label{fig:statemachine}
\end{figure}

\subsection{Fair Scheduling Class Realization (\texttt{kernel/sched/fair.c})}
Handling \texttt{SNT\_REPICK} within \texttt{set\_next\_task\_fair()} requires decoupling timer maintenance from structural entity accounting:
\begin{enumerate}
    \item \textbf{Bypassing Structural State Updates:} On a repick, the task is already active on the CPU. The scheduler skips \texttt{clear\_buddies()}, \texttt{update\_misfit\_status()}, and task group throttle validation.
    \item \textbf{Entity Runtime Refresh Guarantee:} Under EEVDF, before \texttt{pick\_task\_fair()} selects an entity again, it invokes \texttt{update\_curr()} to refresh virtual runtime metrics and slice consumption. Therefore, \texttt{rq->cfs.curr} reflects fresh runtime metrics.
    \item \textbf{Two-Stage Timer Re-Arming:} The repick branch jumps directly to timer re-arming. \texttt{hrtick\_start\_fair()} computes the quantum duration and stores the delay offset; when \texttt{\_\_schedule()} exits and releases runqueue locks, \texttt{hrtick\_schedule\_exit()} programs the physical timer with zero lock contention.
\end{enumerate}

The next listing is abridged pseudocode. It isolates the new repick branch; the complete implementation, including fair-class accounting and task-group handling, is in the merged commit \cite{kaushik2026repick}.

\begin{lstlisting}
static void set_next_task_fair(struct rq *rq, struct task_struct *p, enum snt_e type)
{
    struct sched_entity *se = &p->se;
    struct cfs_rq *cfs_rq = &rq->cfs;
    bool first = (type == SNT_PICK);

    if (type == SNT_REPICK)
        goto repick;

    clear_buddies(cfs_rq, se);
    if (se->on_rq) {
        /* Standard accounting and queue management */
        update_stats_wait_end_fair(cfs_rq, se);
        __dequeue_entity(cfs_rq, se);
        update_load_avg_fair(cfs_rq, se);
    }

    update_misfit_status(p, rq);
    sched_fair_update_stop_tick(rq, p);

repick:
    /*
     * A same-task repick skips put_prev_task_fair(), but
     * pick_task_fair() refreshed the entity hrtick_start_fair() reads
     * before selecting it again. rq->cfs.curr identifies that entity,
     * including under group scheduling hierarchies.
     */
    if (hrtick_enabled_fair(rq))
        hrtick_start_fair(rq, p);
}
\end{lstlisting}

\subsection{Cross-Class Invariance \& Real-Time Safety Analysis}
A critical requirement of our design is ensuring cross-class safety across all scheduling classes in the Linux kernel:

\subsubsection{SCHED\_DEADLINE Safety (\texttt{kernel/sched/deadline.c})}
A naive implementation would restart timers uniformly across all scheduling classes. In \texttt{SCHED\_DEADLINE}, that would use state whose runtime is not refreshed by the same-task picker. Same-task repicks skip \texttt{put\_prev\_task()}, and \texttt{pick\_task\_dl()} has no counterpart to the fair current-runtime update. Restarting \texttt{start\_hrtick\_dl()} on \texttt{SNT\_REPICK} could therefore use stale \texttt{dl\_se->runtime} and defer the next runtime-enforcement point:
\begin{equation}
\Delta t_{\text{armed}} = \frac{\text{dl\_se}\to\text{runtime}_{\text{stale}}}{\text{scaling\_factor}}
\end{equation}
Consequently, \texttt{set\_next\_task\_dl()} explicitly exits immediately upon \texttt{SNT\_REPICK}:

\begin{lstlisting}
static void set_next_task_dl(struct rq *rq, struct task_struct *p, enum snt_e type)
{
    struct sched_dl_entity *dl_se = &p->dl;
    struct dl_rq *dl_rq = &rq->dl;

    if (type == SNT_REPICK)
        return;
    p->se.exec_start = rq_clock_task(rq);
    if (on_dl_rq(&p->dl))
        update_stats_wait_end_dl(dl_rq, dl_se);

    WARN_ON_ONCE(dl_rq->curr);
    dl_rq->curr = dl_se;

    if (type != SNT_PICK)
        return;

    if (rq->donor->sched_class != &dl_sched_class)
        start_hrtick_dl(rq, dl_se);
}
\end{lstlisting}

\subsubsection{SCHED\_RT, SCHED\_IDLE, SCHED\_EXT, and Stop Task}
For \texttt{rt.c}, \texttt{idle.c}, \texttt{stop\_task.c}, and the extensible BPF scheduler class \texttt{ext/ext.c} \cite{linux2026schedext}, \texttt{SNT\_REPICK} returns immediately, preventing redundant accounting and isolating execution benefits strictly to the proportional-share domain.

\section{Evaluation and Scope}

\subsection{Validation Platforms}
We used two complementary validation paths. Platform~A was an 80-core Arm Neoverse-N1 Ampere Altra server at 3.0 GHz with 128 GB DDR4-3200 memory. It was used for functional validation with fair hrtick and \texttt{DELAY\_DEQUEUE} enabled, with \texttt{base\_slice\_ns=3000000}. Smoke tests exercised deadline hrtick paths, and a build with \texttt{CONFIG\_HIGH\_RES\_TIMERS=n} checked the fallback path. These checks establish configuration compatibility, not a standalone quantitative latency result; we therefore do not report local percentile claims without the corresponding trace data and scripts.

Platform~B is the public Intel 0-Day Kernel Test Service result reported for a dual-socket, 192-thread Intel Xeon 6740E (Sierra Forest) system with 256 GB memory \cite{lkp2026rawudp}. The reported job used \texttt{stress-ng} \texttt{rawudp}, 100\% thread count, a 60-second test duration, GCC~14, the performance CPU-frequency governor, the \texttt{x86\_64-rhel-9.4} kernel configuration, and a Debian~13 root filesystem. The service's job specification and reproduction materials are public \cite{lkp2026materials}.

\subsection{Public 0-Day \texttt{rawudp} Comparison}
Table~\ref{tab:xeon_results} transcribes the metrics reported by the 0-Day comparison. The standard-deviation column uses the rounded percent values displayed by the public report for the patched result; the report does not disclose the number of repetitions. Metric names are retained from \texttt{stress-ng}; in particular, \texttt{MB\_recv'd\_per\_sec} is a tool-reported metric and is not independently interpreted as wire bandwidth.

\begin{table}[pos=ht]
\caption{Public Intel 0-Day Kernel Test Service comparison of Linux baseline \texttt{a9b3c75705} and commit \texttt{c72945693b} on \texttt{stress-ng.rawudp} \cite{lkp2026rawudp}.}\label{tab:xeon_results}
\begin{tabular*}{\tblwidth}{@{}LLLLR@{}}
\toprule
Metric & Baseline & Patched & Change & Patched std. dev. \\
\midrule
\texttt{packets\_32\_bytes\_received\_per\_sec} & 395.33 & \textbf{530.63} & \textbf{+34.2\%} & $\pm3\%$ \\
\texttt{ops} & 4,559,098 & \textbf{6,112,029} & \textbf{+34.1\%} & $\pm3\%$ \\
\texttt{ops\_per\_sec} & 76,041 & \textbf{101,916} & \textbf{+34.0\%} & $\pm3\%$ \\
\texttt{MB\_recv'd\_per\_sec} & 0.01 & \textbf{0.02} & \textbf{+100.0\%} & --- \\
Involuntary context switches & 2,611,008 & \textbf{2,999,104} & \textbf{+14.9\%} & $\pm5\%$ \\
User time (s) & 47.70 & \textbf{69.96} & \textbf{+46.7\%} & $\pm8\%$ \\
Voluntary context switches & 20,919,233 & \textbf{29,638,128} & \textbf{+41.7\%} & $\pm2\%$ \\
\bottomrule
\end{tabular*}
\end{table}

The public comparison reports a 34.2\% increase in \texttt{packets\_32\_bytes\_received\_per\_sec}. It is consistent with the repaired hrtick activation path, but it does not isolate a causal mechanism: the available report does not measure socket-buffer occupancy, packet drops, lock contention, cache behavior, or per-task runqueue delay. Those mechanisms require separate scheduler and network instrumentation and are left as future evaluation work.

\subsection{Limitations}
The patch affects the fair scheduling class only when hrticks are enabled. The conditional timing bound in Section~2 does not cover tickless operation, CPUs with a stopped periodic tick, or workloads in which another event invokes scheduling before the next tick. The public \texttt{rawudp} result is one workload on one service configuration; it should not be generalized to end-to-end network throughput, packet loss, or arbitrary parallel workloads. A broader performance study should publish raw traces, experiment scripts, run counts, and confidence intervals for each platform.

\section{Related Work}
Proportional-share scheduling traces its lineage to Generalized Processor Sharing (GPS) \cite{parekh1993gps} and Deficit Round Robin \cite{shreedhar1996drr}. Linux CFS made virtual-runtime scheduling widely deployed, while EEVDF separates eligibility and virtual-deadline selection from the allocation model \cite{lozi2016cfs, stoica1996eevdf, zijlstra2023eevdf}. The present work concerns the activation path after task selection rather than the EEVDF ordering rule itself.

High-resolution timers provide the infrastructure required for short scheduler timing events \cite{gleixner2006hrtimers}. Low-latency RPC systems motivate careful control of scheduling delay \cite{kalia2019fasst}. Kernel documentation describes the extensible BPF scheduling class (\texttt{sched\_ext}) \cite{linux2026schedext}; kernel-bypass operating systems, including IX \cite{belay2014ix} and Arrakis \cite{peter2014arrakis}, pursue different approaches to low-latency networking. To our knowledge, the specific interaction between one-shot fair hrtick expiry and same-task repicks was not documented before the upstream patch \cite{kaushik2026repick}.

\section{Conclusion}
This paper analyzed and repaired a missed fair-hrtick restart after same-task repicks in Linux. The \texttt{SNT\_REPICK} activation type makes that path explicit, restarts the fair hrtick from refreshed state, and leaves deadline hrticks untouched when their runtime state is not refreshed. The implementation is merged as commit \texttt{c72945693b} \cite{kaushik2026repick}. Functional validation covers fair hrtick, deadline hrtick, and the high-resolution-timer fallback; a public 0-Day comparison reports a 34.2\% improvement in one \texttt{stress-ng.rawudp} metric \cite{lkp2026rawudp}. The result motivates a broader study with per-task traces and repeated, artifact-backed measurements across workloads and tick configurations.

\section*{Acknowledgements}
The author extends sincere gratitude to Peter Zijlstra (Linux Kernel Scheduler Maintainer, Intel) for valuable technical discussions, design suggestions, and upstream review and integration of the core scheduling patch into the Linux kernel tree. The author also thanks Zhan Xusheng for patch review on the Linux Kernel Mailing List, and the Intel 0-Day Kernel Test Robot engineering team for automated hardware performance validation.

\printcredits

\section*{Declaration of Competing Interests}
The author declares that they have no known competing financial interests or personal relationships that could have appeared to influence the work reported in this paper.

\section*{Declaration of Generative AI and AI-Assisted Technologies}
During the preparation of this work, the author used generative AI tools strictly for grammar, style checking, and LaTeX typesetting alignment against the Elsevier Journal of Parallel and Distributed Computing author guidelines. After using this service, the author reviewed and edited the content as needed and takes full responsibility for the content of the published article.

\section*{Data Availability Statement}
The Linux kernel patch, public performance report, and associated reproduction materials are accessible from the following public records:
\begin{itemize}
    \item \textbf{Mainline Git Commit:} \url{https://git.kernel.org/pub/scm/linux/kernel/git/torvalds/linux.git/commit/?id=c72945693b90423f1b46ac2dbc8749c2b7804fc8}
    \item \textbf{Upstream Repository:} \url{https://git.kernel.org/pub/scm/linux/kernel/git/torvalds/linux.git}
    \item \textbf{LKML v4 Submission:} \url{https://lore.kernel.org/all/20260917-sched-fair-hrtick-restart-v4-1-4dd1414da81a@gentwo.org/}
    \item \textbf{0-Day Benchmark Result:} \url{https://lore.kernel.org/lkml/202610271134.d6bf1e10-lkp@intel.com/}
    \item \textbf{0-Day Reproduction Materials:} \url{https://download.01.org/0day-ci/archive/20261027/202610271134.d6bf1e10-lkp@intel.com}
\end{itemize}


\begin{thebibliography}{99}

\bibitem{parekh1993gps}
A. K. Parekh and R. G. Gallager,
``A generalized processor sharing approach to flow control in integrated services networks: The single-node case,''
\emph{IEEE/ACM Transactions on Networking}, vol. 1, no. 3, pp. 344--357, 1993.
\href{https://doi.org/10.1109/90.234856}{https://doi.org/10.1109/90.234856}

\bibitem{stoica1996eevdf}
I. Stoica, H. Abdel-Wahab, K. Jeffay, S. K. Baruah, J. E. Gehrke, and C. G. Plaxton,
``A proportional share resource allocation algorithm for real-time, time-shared systems,''
in \emph{Proceedings of the 17th IEEE Real-Time Systems Symposium}, pp. 288--299, 1996.
\href{https://doi.org/10.1109/REAL.1996.563725}{https://doi.org/10.1109/REAL.1996.563725}

\bibitem{shreedhar1996drr}
M. Shreedhar and G. Varghese,
``Efficient fair queuing using deficit round-robin,''
\emph{IEEE/ACM Transactions on Networking}, vol. 4, no. 3, pp. 375--385, 1996.
\href{https://doi.org/10.1109/90.502236}{https://doi.org/10.1109/90.502236}

\bibitem{lozi2016cfs}
J.-P. Lozi, B. Lepers, J. Funston, F. Gaud, V. Qu\'ema, and A. Fedorova,
``The Linux scheduler: A decade of wasted cores,''
in \emph{Proceedings of the Eleventh European Conference on Computer Systems}, pp. 1--16, 2016.
\href{https://doi.org/10.1145/2901318.2901326}{https://doi.org/10.1145/2901318.2901326}

\bibitem{corbet2023eevdf}
J. Corbet, ``An EEVDF CPU scheduler for Linux,'' \emph{LWN.net}, Mar. 2023.
\url{https://lwn.net/Articles/925371/}

\bibitem{zijlstra2023eevdf}
P. Zijlstra, ``[PATCH 00/15] sched/fair: Complete EEVDF implementation,''
\emph{Linux Kernel Mailing List}, May 2023.
\url{https://lore.kernel.org/lkml/20230531124603.650428985@infradead.org/}

\bibitem{kaushik2026repick}
S. Kaushik, ``sched: Restart fair hrtick after same-task repicks,''
\emph{Linux kernel source tree}, commit c72945693b90423f1b46ac2dbc8749c2b7804fc8, Sep. 2026.
\url{https://git.kernel.org/pub/scm/linux/kernel/git/torvalds/linux.git/commit/?id=c72945693b90423f1b46ac2dbc8749c2b7804fc8}

\bibitem{lkp2026rawudp}
Kernel Test Robot, ``[linus:master] [sched] c72945693b: \texttt{stress-ng.rawudp.packets\_32\_bytes\_received\_per\_sec} 34.2\% improvement,''
\emph{Linux Kernel Mailing List}, Oct. 2026.
\url{https://lore.kernel.org/lkml/202610271134.d6bf1e10-lkp@intel.com/}

\bibitem{lkp2026materials}
Kernel Test Robot, ``Reproduction materials for stress-ng rawudp result 202610271134.d6bf1e10-lkp@intel.com,'' Oct. 2026.
\url{https://download.01.org/0day-ci/archive/20261027/202610271134.d6bf1e10-lkp@intel.com}

\bibitem{gleixner2006hrtimers}
T. Gleixner and D. Niehaus,
``Hrtimers and beyond: Transforming the Linux time subsystems,''
in \emph{Proceedings of the Ottawa Linux Symposium}, pp. 333--346, 2006.
\url{https://www.kernel.org/doc/ols/2006/ols2006v1-pages-333-346.pdf}

\bibitem{linux2026schedext}
The Linux Kernel documentation project, ``Extensible Scheduler Class,'' 2026.
\url{https://docs.kernel.org/scheduler/sched-ext.html}

\bibitem{kalia2019fasst}
A. Kalia, M. Kaminsky, and D. G. Andersen,
``Datacenter RPCs can be general and fast,''
in \emph{16th USENIX Symposium on Networked Systems Design and Implementation (NSDI 19)}, pp. 1--16, 2019.
\url{https://www.usenix.org/conference/nsdi19/presentation/kalia}

\bibitem{belay2014ix}
A. Belay, G. Prekas, A. Klimovic, S. Grossman, C. Kozyrakis, and E. Bugnion,
``IX: A protected dataplane operating system for high throughput and low latency,''
in \emph{11th USENIX Symposium on Operating Systems Design and Implementation (OSDI 14)}, pp. 49--65, 2014.
\url{https://www.usenix.org/conference/osdi14/technical-sessions/presentation/belay}

\bibitem{peter2014arrakis}
S. Peter, J. Li, I. Zhang, D. R. K. Ports, D. Woos, A. Krishnamurthy, T. Anderson, and T. Roscoe,
``Arrakis: The operating system is the control plane,''
in \emph{11th USENIX Symposium on Operating Systems Design and Implementation (OSDI 14)}, pp. 1--16, 2014.
\url{https://www.usenix.org/conference/osdi14/technical-sessions/presentation/peter}

\end{thebibliography}
\end{document}